\documentclass[prl,aps,amsmath,amssymb,amsfonts,floatfix,groupedaddress,
showpacs,twocolumn]{revtex4}
\usepackage{graphicx}
\usepackage{epsfig}
\usepackage[dvips]{color}
\usepackage{dcolumn}
\newcommand{\bR}{\mathbf{R}}
\newcommand{\bK}{\mathbf{K}}

\def\beq{\begin{equation}}
\def\eeq{\end{equation}}
\def\be{\begin{equation}}
\def\ee{\end{equation}}

\def\t{\mbox{tr}\,}
\def\cG0{{\cal G}_0}
\def\cG{{\cal G}}

\def\a{\alpha}

\def\uc2{$U_{c2}$}
\def\uc1{$U_{c1}$}

\def\fd{f^\dagger}

\def\ket{\rangle}

\def\bran{\langle n|}

\def\phiAn{\phi_{An}^{\hfill}}

\def\bavs3{BaVS$_3$}

\def\t2g{$t_{2g}$}
\def\eeg{$e_{g}$}
\def\a1g{$a_{1g}$}

\begin{document}
\title{Correlation effects on the doped triangular lattice in view of the 
physics of sodium-rich Na$_x$CoO$_2$}
\author{Frank Lechermann}
\affiliation{I. Institut f{\"u}r Theoretische Physik, 
Universit{\"a}t Hamburg, D-20355 Hamburg, Germany}

\begin{abstract}
The peculiar correlation effects on the triangular lattice are studied by means
of the rotationally invariant slave boson method in a cellular cluster 
approach. Hence nonlocal correlations are included in a short-range regime.
Their impact for the single-band Hubbard model is studied at half filling, 
i.e., on the Mott transition, and with doping. Using the realistic band 
structure of Na$_x$CoO$_2$, we may also shed light on the cobaltate physics for
$x$$\ge$$1/3$, with the in-plane transition from antiferromagnetic 
tendencies towards the onset of ferromagnetism for a finite Hubbard $U$.
\end{abstract}

\maketitle
The transition-metal oxide Na$_x$CoO$_2$, consisting of CoO$_2$ layers and Na 
ions inbetween, is one of the most fascinating examples of a doped correlated 
electron system on the triangular lattice. Albeit evidently the bare 
CoO$_2$ ($x$=0) compound is metallic~\cite{vau07}, surprisingly, the effect of 
strong correlation appears to be severe close to the band-insulating ($x$=1) 
limit. Several experimental studies have revealed a rather rich phase diagram 
with highlighted physical properties such as superconductivity close to $x$=0.3 
(when intercalated with H$_2$O)~\cite{tak03}, Pauli(Curie-Weiss)-like 
metallicity for $x$$<$0.5($>$0.5)~\cite{foo04}, in-plane antiferromagnetic 
(AFM) order at $x$=0.5~\cite{men05} , large thermopower around 
$0.71$$<$$x$$<$$0.84$~\cite{wan03}, charge disproportionation for 
$x$$>$0.5~\cite{muk05} and in-plane ferromagnetic (FM) order for 
$0.75$$<$$x$$<$$0.9$~\cite{boo04,bay05}. 
Due to the sizable (\t2g,\eeg) crystal-field (CF) splitting, the Co ion is 
expected to be in a low-spin state, with $x$ controlling the 
residual occupation of the \t2g manifold. Hence Co$^{4+}$ ($S$=1/2) for $x$=0 
and Co$^{3+}$ ($S$=0) for $x$=1. Yet the respective \t2g fillings and apparent
Fermi surface (FS) for smaller $x$ are still a matter of debate~\cite{mar07}. 
Though calculations based on the local density approximation (LDA) yield an
\t2g-internal \a1g-$e_g'$ CF splitting of -0.1 eV~\cite{lecproc,pil08}, only 
the \a1g-like bands are expected to be partially depleted for larger 
$x$. Those should form a single hexagonal FS sheet that is hole-like, 
i.e., the hopping $t$ within a nearest-neighbor (NN) tight-binding (TB) model 
is negative.

Concerning magnetism, LDA suggests FM order for already small $x$, although the 
AFM state is very close in energy~\cite{sin00,joh05}. Dynamical mean-field 
theory (DMFT) studies for the NN-TB Hubbard model on a triangular 
lattice~\cite{mer06,ary06} show an instability towards FM order only for 
$t$$>$0. Gao {\sl et al.}~\cite{gao07} described the appearance of a 
renormalized Stoner instability in an infinite-$U$ Gutzwiller treatment of a 
third NN-TB model for Na$_x$CoO$_2$ (with the NN $t$$<$0) at $x$$\sim$0.67. 
Recently, a finite-$U$ LDA+Gutzwiller approach obtained intralayer FM order at 
larger $x$~\cite{wan08}, but contrary to experiment~\cite{vau07} finds full FM 
order at low doping. There are cluster approaches to the Hubbard model on the 
triangular lattice (e.g.~\cite{kyu07,sah08}) but without incorporating the 
detailed electronic structure of Na$_x$CoO$_2$.

\begin{figure}[t]
\includegraphics*[width=3.5cm]{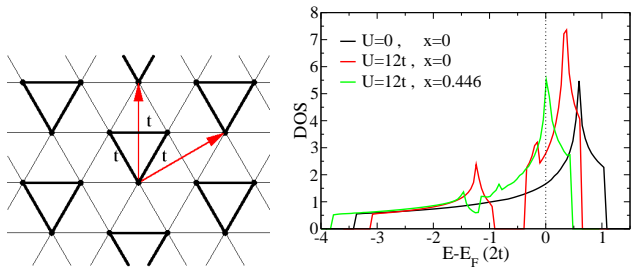}\hspace{0.2cm}
\includegraphics*[width=4.5cm]{qp-dos.tri.eps}
\caption{(color online) 3-site cluster approach on the triangular lattice for
isotropic NN hopping $t$. Left: lattice tiling, right: QP DOS for different
interaction strength and doping.
\label{trilat}}
\end{figure}

In the present work, realistic sodium cobaltate is investigated at larger Na 
doping, finite $U$ and by explicitly including NN correlation effects via
the cellular cluster scheme. The puzzling change from AFM tendencies at
small $x$ towards the onset of intralayer FM order may be described within a 
Hubbard-like model using LDA dispersions and moderate $U$. The recently
generalized rotationally invariant slave boson mean-field theory~\cite{lec07} 
(RISB) is applied to the problem~\footnote{A different slave spin cluster mean 
field theory was presented recently in a similar context by S.~R.~Hassan and 
L.~de'~Medici, arXiv:0805.3550 (2008).}, thereby tiling the lattice into 
NN triangles (see Fig.~\ref{trilat}). To benchmark this approach in the present
setup, we first studied the NN-TB single-band Hubbard model, written in 
the cellular cluster scheme as $H$=$H(\bK)+H_{\alpha}$, i.e.,
\begin{equation}
H=\sum_{\bK ij\sigma}\varepsilon_{ij\sigma\sigma}^{\bK}
c^{\dagger}_{\bK i\sigma}c^{\hfill}_{\bK j\sigma}
-t\sum_{\alpha ij\sigma}c^{\dagger}_{\alpha i\sigma}c^{\hfill}_{\alpha j\sigma}
+U\sum_{\alpha i}n_{\alpha i\uparrow}n_{\alpha i\downarrow}\,\,,
\label{cellhub}
\end{equation}
where $\alpha$ marks the cluster, $ij$ are site indices on the cluster, 
$\sigma$ denotes the spin $(\uparrow,\downarrow)$ and $\bK$ is
the cluster wave vector. Note that for the cluster dispersion
$\sum_{\bK}\varepsilon_{ij\sigma\sigma}^{\bK}$=0 holds and the intracluster
hopping is taken care of by the second term in (\ref{cellhub}). Within RISB
the electron operator $c_{i\sigma}$ is represented as 
$\underline{c}_{i\sigma}$=$\hat{R}[\phi]^{\sigma\sigma'}_{ij}f_{j\sigma'}$, 
where $R$ is a non-diagonal transformation operator that relates the physical 
operator to the quasiparticle (QP) operator $f_{i\sigma}$. The transformation 
$\hat{R}$ is written in terms of slave bosons $\{\phi_{An}\}$ with two 
indices, namely $A$ for the physical-electron state and $n$ for the QP Fock 
state. It follows that the kinetic part of (\ref{cellhub}) is expressed via 
the QP operators with renormalized dispersions and the operator character of
the local part is carried solely by the slave bosons~\cite{lec07}.
Two constraints, namely
$\sum_{An}\phi^\dagger_{An}\phiAn$=1 and 
$\sum_{Ann'} \phi^\dagger_{An'}\phiAn\bran\fd_{i\sigma} f_{j\sigma'}^{\hfill}|n'\ket$=$f_{i\sigma}^{\dagger}\,f_{j\sigma'}^{\hfill}$,
to select the physical states are imposed through a set of Lagrange multipliers 
$\{\lambda\}$. A saddle-point solution is obtained by condensing the bosons 
and extremalizing the corresponding free energy 
$F(\{\phi\},\{\lambda\})$~\footnote{Gaussian smearing of $\sim$1 meV is used
for the K point integration, introducing a small effective temperature.}.
The nonlocal QP weight matrix at saddle point reads 
${\bf Z}$=$\bR\bR^{\dagger}$.

\begin{figure}[b]
\includegraphics*[width=8.25cm]{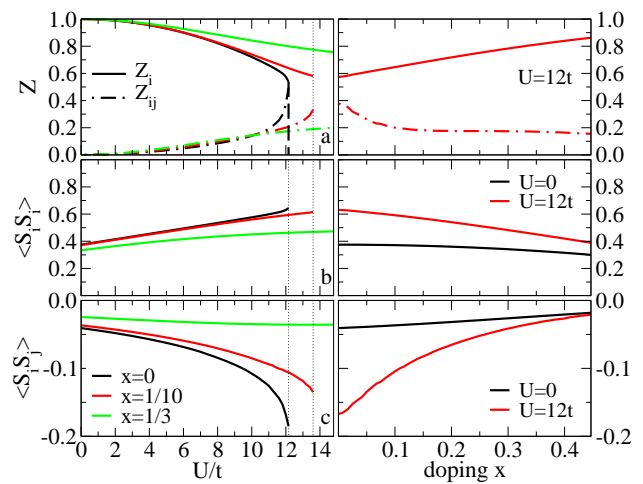}
\caption{(color online) Observables for the NN-TB Hubbard model ($t$$>$0) on the
triangular lattice. Left: $U$ dependence for (a) QP weights, 
(b) onsite spin correlation and (c) 
NN spin correlation. Right: doping dependence of the same observables. The
$Z_{ij}$ are multiplied by 5 for better visibility.
\label{mtri}}
\end{figure}

In the following, we study the paramagnetic (PM) phase where $x$ denotes the 
electron doping normalized to a single orbital. The QP density of states (DOS) 
of the noninteracting model (bandwidth $W$=9$t$), along with comparative 
interacting cases, is shown in Fig.~\ref{trilat}. As displayed in 
Fig.~\ref{mtri}, the RISB treatment for $x$=0 reveals a first-order Mott 
transition at $U_c$$\sim$$12.2t$=1.36$W$, in agreement with 
values from exact diagonalization~\cite{cap01} ($\sim$12.1$t$), cellular 
DMFT~\cite{kyu07} ($\sim$10.5$t$) and variational cluster 
approximation~\cite{sah08} ($\sim$12$t$). Upon doping, the metallic state may be
extended to larger $U$, although a breakdown of the conventional PM phase is 
found for small $x$ at $U$$>$$U_c$. The magnitude of the offdiagonal QP weights 
$Z_{ij}$ as well as of the NN spin correlation
$\langle {\bf S}_i{\bf S}_j\rangle$ are enhanced close to the Mott transition. 
Note that $\langle {\bf S}_i{\bf S}_j\rangle$ is negative, i.e., of AFM 
character. Close to half filling, the $Z_{ij}$ depend strongly on $x$ and the 
onsite spin correlation is substantially enhanced for $U$$\not=$0. The AFM 
strength of the intersite $\langle {\bf S}_i{\bf S}_j\rangle$ is decreasing 
with $x$, but remains slightly above the $U$=0 case in the (meta)stability range
of the PM phase. Note that the {\sl global} FM instability was located around 
$x$$\sim$0.35~\cite{mer06}.

\begin{figure}[t]
\includegraphics*[width=8cm]{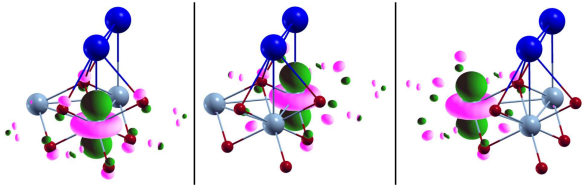}
\caption{(color online) From left to right: $a_{1g}$-like WFs in Na$_x$CoO$_2$
unit cell for $x$=2/3 on Co(1), Co(2) and Co(3). Large blue (dark) balls 
denote Na ions, small red (gray) ball mark O ions.
\label{nadop1}}
\end{figure}

For a realistic description of Na$_x$CoO$_2$ we first performed LDA calculations 
at different stoichiometries. A supercell involving the NN Co triangle of a 
given CoO$_2$ layer served as the base structure. Decorating the 
latter with additional Na ions {\sl above} the Co sites yields the 
dopings $x$=0,1/3,2/3,1. No sodium ions above/below oxygen 
positions are considered, which should be adequate for the overall behavior  
with doping. A minimal model for the band structure is derived by representing 
the low-energy states by three maximally-localized Wannier 
functions~\cite{mar97} (WFs) of $a_{1g}$ type (see Fig.~\ref{nadop1}). Thus the 
full dispersions ($W$$\sim$1.2 eV) are approximated through a downfolding 
procedure~\cite{sou01} via three $a_{1g}$-like bands. 
Allowing for the essential behavior with $x$, a doping dependent hamiltonian is 
obtained through a linear interpolation 
$H_{\rm LDA}(\bK,x)$=$xH_{\rm LDA}^{x_a}(\bK)$$+$$(1$$-$$x)H_{\rm LDA}^{x_b}(\bK)$,
where $x_a$,$x_b$ are the neighboring LDA-treated dopings. Note that the K 
points are generally on a 3-dimensional (3D) mesh, describing the full band 
dispersions. In the full hamiltonian (\ref{cellhub}) the kinetic term is now 
given by $H_{\rm LDA}(\bK,x)$ {\sl without} its onsite (cluster) term $t_{ij}$. 
The latter replaces the model $t$ in the ``onsite'' quadratic term of 
(\ref{cellhub}), incorporating both, the intersite hoppings on the Co
cluster and the CF-terms on the individual ions. Be aware that the level of
frustration is diminished via the symmetry-breaking Na occupations for 
intermediate $x$.

\begin{figure}[t]
\includegraphics*[width=8.6cm]{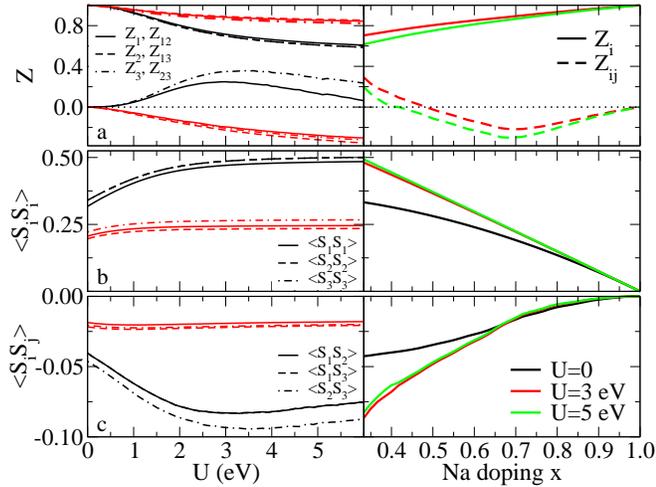}
\caption{(color online) Observables for Na$_x$CoO$_2$.
Left: (a) site-resolved QP weights, (b) onsite- and (c) NN spin correlations 
for fixed doping. $x$=1/3: (black/dark) lines, $x$=2/3: (red/gray) lines.
The offdiagonal $Z_{ij}$ go to zero at $U$=0. Co
sites are numbered with Na filling: '1' has Na on top at $x$=1/3, '2'
also at $x$=2/3. Right: doping dependence of the averaged 
observables. All $Z_{ij}$ are multiplied by 50 for better visibility.
\label{nadop2}}
\end{figure}
Contrary to the former model study, the realistic NN hoppings are now 
negative~\cite{mer06}. Their absolute value is more or less linearly decreasing 
with $x$, in average from $0.21$ eV ($x$=0) to $0.11$ eV ($x$=1). Any further 
explicit exchange couplings $J$ are neglected. Integrating out the O states 
should render such couplings necessary (e.g., (A)FM exchange between the Co 
ions). Yet a rigorous computation for metals is hard~\cite{joh05} and 
experiments revealed~\cite{bay05} that the values for a {\sl pure} spin 
hamiltonian are small. Calculations including Heisenberg-like terms in $H$ 
resulted in no qualitative change for reasonable values of $J$ (see 
footnote on the spin susceptibility). No long-range Coulomb potential is 
included, hence explicit charge-disproportionated states do not appear (though
site occupations surely differ). We concentrate the discussion on the 
phases for $x$$\ge$1/3, since neglecting the remaining $e_g'$
orbitals at small $x$ may not be improper. In fact, when performing the RISB 
calculation at $x$=0, the Mott transition sets in at $U_c$$\sim$ 3.1 eV. But 
this value is surely too small because of our orbital restrictions. 

Figure~\ref{nadop2} shows RISB results for $x$=1/3,2/3 and with varying Na 
content for reasonable ranges/values of the interaction strength $U$. The 
resulting moderate size of the diagonal QP weight is in good agreement with 
experiment. Note that for $x$=1/3 the offdiagonal $Z_{ij}$ between sites on the 
triangle is positive with a maximum at $U$$\sim$3 eV. For $x$=2/3 its 
value is negative and still increasing in magnitude in the studied $U$ range. 
The sign change implies that the hierachy in the correlation strength within the
bonding-antibonding cluster orbitals is reversed. The onsite
spin correlations are saturating for large $U$, while the intersite 
$\langle {\bf S}_i{\bf S}_j\rangle$ are always negative but decrease in 
magnitude in the latter regime. For $x$=1/3 the NN (spin) correlations are 
enhanced in magnitude between Co sites without Na on top. Due to CF effects,
the orbital filling on a Co site tends generally to increase when placing Na on 
top. For the doping study, $U$=3 eV and $U$=5 eV were used. There the 
intersite $Z_{ij}$ changes sign at $x$$\lesssim$0.5 and has a minimum close to 
$x$=2/3. As expected, the local spin correlations are strongly increased 
for finite $U$ close to half filling~\cite{kyu07}. The AFM character of the NN 
spin correlations is strengthened for intermediate $x$ compared to the $U$=0 
case. However interestingly, there is a crossover at $x$$\sim$2/3 where the AFM 
magnitude for $U$$\not=$0 becomes lower than in the latter case. Thus the AFM 
tendencies are surpressed by electronic correlations for large $x$, where 
the superexchange is diminished. This seems reminiscent of the Nagaoka 
mechanism~\cite{nag66} for the infinite-$U$ Hubbard model on the square lattice,
leading to FM order. But the Nagaoka state is not a ground state on the 
triangular lattice for a single hole~\cite{hae05}. 

\begin{figure}[t]
\includegraphics*[width=8.7cm]{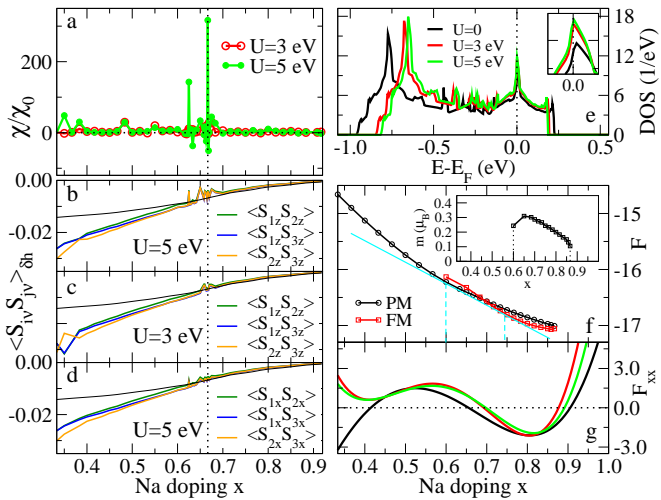}\hspace{0.3cm}
\caption{(color online) Ordering tendencies and phase stability in
Na$_x$CoO$_2$. (a) normalized magnetic susceptibility, (b-d) component-resolved
NN spin correlation in the small field $\delta h$. Solid black line:
$\langle S_{i\nu}S_{j\nu}\rangle_{\delta h}$ for $U$=0. (e) QP DOS for
$x$=2/3. (f) RISB free energies for PM and FM phase of the
single layer for $U$=5 eV. Straight cyan (lightgray) line: common 
tangent. Inset: net magnetic moment per Co ion in the FM phase.
(g) Second derivative of the full PM free energy.
\label{nasus}}
\end{figure}

To elucidate the problem of magnetic instabilities, a residual magnetic field 
${\bf H}_{\rm f}$=$\delta h\,{\bf e}_z$ is applied in the PM phase. Therefrom 
the uniform magnetic susceptibility is determined as 
$\chi$$\equiv$$\partial M$/$\partial H_{\rm f}$$\approx$$\delta M$/$\delta h$, 
where $\delta M$ is the residual magnetization induced by $\delta h$. 
Figure~\ref{nasus}a shows the normalized response $\chi/\chi_0$, where $\chi_0$
is the susceptibility for $U$=0~\footnote{A magnetic response of the cobaltate
model for $U$=0 may surely be achieved via an explicit exchange term.}. It is 
seen that there is strong response for $x$$\sim$2/3 and $U$=5 eV, similar to 
results 
in the infinite-$U$ limit~\cite{gao07}. Additionally, there is already a 
precursive regime for $x$$\gtrsim$0.62 where FM ordering tendencies show up.
This coincides with recent NMR measurements by Lang {\sl et al.}, who find a
crossover from AFM to FM correlations at $x$=0.63-0.65~\cite{lan08}. From the
site- and component resolved intersite spin correlations in the applied field,
the dominant response of $\langle S_{iz}S_{jz}\rangle$ in the doping regime 
0.62$\lesssim$$x$$\lesssim$0.7 is again obvious. Note that also $U$=3 eV 
exhibits minor FM ordering tendencies there. Interestingly, for $x$$\sim$0.35 one
may observe intralayer AFM response for both values of $U$. An AFM ordering 
signal is identified through locally favorable spin antialignment between 
Co(1)-Co(2) and Co(1)-Co(3), whereas Co(2)-Co(3) favor spin alignment. Remember 
that for $x$=1/3 the Na ion is above Co(1), thus the Co differentiation via 
neighboring Na ions obviously triggers the AFM tendencies. The outer-field 
induced AFM response is in line with the observation that such 
a field can lift the effects of kinetic-energy frustration on a triangular
cluster~\cite{bar91}. The fact that no clear long-range order is visible for 
larger $x$ (i.e., negative $\chi/\chi_0$) may be explained by the present 3D 
model that cannot stabilize the A-type AFM phase (no proper interlayer 
resolution). Yet it is possible to stabilize FM order within a single layer by 
using only the 2D-projected LDA dispersion. Figure~\ref{nasus}f shows the 
tie-line construction between the PM and FM phase in this case. A stable 
{\sl homogeneous} FM order is revealed for $x$$\gtrsim$0.74, with magnetic
moments close to experimental values. The onset of the {\sl heterogeneous} phase
at $x$$\sim$0.6 matches the begin of FM tendencies in the full 3D model. 
In this context, hints for a spin-liquid ground state at 
0.71$\le$$x$$<$0.75~\cite{bal08} are rather interesting. Either 2D phase
becomes hard to stabilize numerically for $x$$\gtrsim$0.87.
Though at $x$=2/3 the Fermi level is located in the upper maximum peak of the
QP DOS (see Fig.~\ref{nadop2}e), our study reveals that a Stoner 
instability can not be the sole origin for magnetic ordering. The nonmonotonic 
behavior of $Z_{ij}$, the subtle change of the frustration level with $x$, the 
precursive FM regime, the PM-FM phase competition as well as the breakdown of 
the 2D phases at very large doping are all pointing
towards nonlocal correlations as an additional important ingredient. As a side 
effect, plotting $\partial^2F/\partial x^2$ (Fig.~\ref{nadop2}g) shows 
that the present 3D-PM phase becomes unstable with respect to doping 
for $x$$\gtrsim$0.7, while $U$ stabilizes the phase for $x$$<$0.5.
Phase separation is experimentally indeed observed at large $x$ 
(e.g.~\cite{lee06} and references therein).

In conclusion, it was shown that the RISB formalism may describe the
essential physics of the NN-TB Hubbard model on the triangular lattice, 
including the first-order Mott transition and the Fermi-liquid upon 
doping. The magnetic behavior with doping in Na$_x$CoO$_2$ can be understood 
with a cluster Hubbard model using LDA dispersions. In very good  
quantitative agreement with experimental data, the change from AFM to FM 
tendencies with a final first-order transition in $x$ into an in-plane FM
phase was revealed. For the correlated 
physics with intralayer FM tendencies starting at $x$$\gtrsim$0.62 and 2D-FM 
order for 0.74$<$$x$$<$0.87, an interaction strength $U$$>$3 eV is sufficient. 
Since then $U/W$$\gtrsim$2.5 and the strongly-correlated regime is already 
reached, without implying that $x$=0 is Mott insulating 
(due to the probably increased orbital fluctuations there). Further 
studies, also at small doping, including long-range and interlayer correlations 
are needed to clarify more details.

\acknowledgements
The author is indebted to M.~Potthoff, I.~I.~Mazin and P.~S.~Cornaglia for 
helpful discussions. This work was supported by the SFB 668.

\bibliographystyle{apsrev}
\bibliography{bibextra}

\end{document}